\documentclass{aa}
\usepackage{graphicx}
\usepackage[varg]{txfonts}
\usepackage{natbib}
\bibpunct{(}{)}{;}{a}{}{,} 
\begin{document}

\title{Dust and gas density evolution at a radial pressure bump in protoplanetary disks}
   \author{Tetsuo Taki\inst{1,2} \and
          Masaki Fujimoto\inst{3,4} \and
          Shigeru Ida\inst{4}
          }

\institute{Center for Computational Astrophysics, National Astronomical Observatory of Japan, Osawa, Mitaka, Tokyo 181-8588, Japan
           \email{tetsuo.taki@nao.ac.jp} \and
	   Department of Earth and Planetary Science, Tokyo Institute of Technology, Meguro-ku, Tokyo 152-8550, Japan \and
          Institute of Space and Astronomical Science, Japan Aerospace Exploration Agency, Yoshinodai 3-1-1, Sagamihara, Kanagawa, Japan
              \and
              Earth-Life Science Institute, Tokyo Institute of Technology, Meguro-ku, Tokyo 152-8550, Japan
             }

%
\abstract{
We investigate the
simultaneous evolution of dust and gas density profiles at a radial pressure bump located in a protoplanetary disk.
If
dust particles are treated as test particles, a radial pressure bump traps dust particles that drift radially inward.
As the dust particles become more
concentrated at the gas pressure bump, however, the drag force from dust to gas (back-reaction),
which is ignored in a test-particle approach,
deforms the pressure bump.
We
find that the pressure bump is completely deformed by the back-reaction
when the dust-to-gas mass ratio reaches $\sim 1$ for a slower bump restoration.
The direct gravitational instability of dust particles is inhibited by the bump destruction.
In the dust-enriched region,
the radial pressure support becomes $\sim 10-100$ times lower than the global value set initially.
Although the pressure bump is a favorable place for streaming instability (SI), the flattened pressure gradient inhibits SI from forming large particle clumps corresponding to $100-1000$ km sized bodies, which has been previously proposed.
If SI occurs there, the dust clumps formed would be $10-100$ times smaller, that is, of about $1 - 100$ km.
}
\maketitle
\authorrunning{T. Taki et al.}
%
\section{Introduction}
The conventional belief is that planetesimals are the
building blocks of planets.
However, the process of
dust grains that grow into planetesimals in a protoplanetary disk is
poorly understood.
There are many theoretical difficulties in understanding the formation process of planetesimals.

The radial drift barrier is one of the most serious problems to understand in planetesimal formation.
The disk is at hydrostatic equilibrium, in which the pressure and the centrifugal force balances the gravity of the central star.
The pressure gradient causes the disk to rotate at a sub-Keplerian velocity. 
Since the dust particles tend to rotate at a Keplerian velocity, they feel a headwind from the gas flow.
As a result, they lose their angular momentum through gas drag and drift inward \citep{wipple72,1976PThPh..56.1756A,1977MNRAS.180...57W}.
Because the dust growth timescale is generally significantly longer than the radial migration timescale, it is likely that the dust particles are lost from the disk.

Many ideas have been proposed
to bypass the radial drift barrier.
A straightforward solution is that planetesimals form
rapidly before dust grains drift significantly.
It has recently been proposed that the fluffy aggregates with a low bulk density grow more quickly and form icy planetesimals when they enter the Stokes regime \citep{2012ApJ...752..106O, 2013A&A...554A...4K, 2013A&A...557L...4K}.
Another mechanism for rapid planetesimal formation is
self-gravitational collapse to form massive clumps in dust-rich regions (the so-called streaming instability (SI); \citet{2005ApJ...620..459Y, 2007ApJ...662..613Y, 2007ApJ...662..627J}) and the following break-up of the massive clumps into smaller pieces.
The SI is induced
by inward radial drift of solid material
that is due to gas pressure support.
The instability modulates
the spatial density pattern of solid material and forms dense particle clumps.
When concentrated enough, these
clumps may contract
and become planetesimals as a result of self-gravity.
We note that the SI does not work for small dust grains.
The growth rate of SI is at its greatest when the particles are frictionally coupled with the gas on an orbital timescale.
For instance, particles with a radius of $30$ cm are most favorable for SI growth at the orbital distance $r = 5$ AU in the minimum mass solar nebula (MMSN; \citet{1985prpl.conf.1100H}) model \citep{2011EM&P..108...39J}.
For SI to be triggered, the particles must be decimeter-sized bodies in the inner part of protoplanetary disks.

The migration trap is another idea put forward to bypass
the radial drift barrier.
When a protoplanetary disk has a local pressure maximum, called a radial pressure bump, solid materials are trapped at the point where the pressure is radially maximized \citep{wipple72,2003ApJ...583..996H,2003ApJ...598.1301H}.
Radial pressure bumps are believed to form in various
parts of protoplanetary disks, such as at the disk inner edge, at the dead zone inner edge, and at the snow line \citep{Johansen2014}.
From
recent submillimeter observations, it has been discovered that
some protoplanetary disks have an asymmetrical
dust profile \citep{Casassus2013,Perez2014}, and some of these might possess
a radial pressure bump \citep{2013Sci...340.1199V}.

A typical dust-to-gas mass ratio in a protoplanetary disk is $\sim 0.01$.
Since the back-reaction
is proportional to the dust-to-gas mass ratio, neglecting the back-reaction
is a good approximation for dust and gas motion in a general situation.
\citet{2003ApJ...583..996H,2003ApJ...598.1301H} considered the disk gas, and thus a pressure bump, to be
a steady structure.
When a pressure bump accumulates dust particles, however,
the back-reaction from particles to gas becomes stronger (called the dust-dominated phase; \citet{1981Icar...45..517N,1986Icar...67..375N}),
 and eventually, the back-reaction alters the gas
velocity and density distribution of the pressure bump \citep{2012ApJ...747...11K}, which affects its
capability of trapping dust particles.
It is important to take the back-reaction in the simulations of dust accumulation at the pressure bump into account.

\cite{2009ApJ...691.1697K,2010ApJ...714.1155K,2012ApJ...747...11K}
partially confirmed the effects of the back-reaction.
They investigated the inhomogeneous growth of a magneto-rotational instability (MRI) in a local area embedded in a protoplanetary disk.
They found that the inhomogeneous growth of an MRI formed a quasi-steady radial pressure bump and meter-sized boulders were trapped in it.
They found that the highest dust density was significantly lower with the back-reaction than without it.
The back-reaction of dust onto gas,
when included, brought
the gas azimuthal velocity closer to
Keplerian rotation in the dust-dense region, which was not favorable for a subsequent concentration of dust particles.
Their simulations, however, were computationally expensive and included
many complex settings.
They did not follow the long-term evolution of the dust and gas distributions.
In particular, because they also included MRI, the amount of gas surface density evolution that was contributed to by the back-reaction was not clear.

In this paper, we investigate the simultaneous evolution of the dust and gas density profile at a simplified radial pressure bump.
For simplicity, we treat the disk gas as if it were
in the (non-magnetized) hydrodynamic fluid.
In addition, we
set the pressure bump as an initial condition.
The initial pressure balance is maintained by the Coriolis force from azimuthal velocity.
Then the pressure bump is steady in the absence of a back-reaction from dust onto gas,
even without the source that
produced the bump (e.g., MRI zonal flows, edges of the MRI dead zone, a snow line, etc.). \citet{2009ApJ...691.1697K,2010ApJ...714.1155K,2012ApJ...747...11K}
assumed that the MRI turbulence was due to a
radially inhomogeneous external magnetic field and that the MRI does not occur anymore after a
bump structure with local rigid rotation is established.
The dust and gas motions are mutually coupled by the drag forces.
The dust particles tend to accumulate at the point of highest pressure,
while the dust accumulation alters the gas pressure profile.
Since the altered gas profile also affects the dust motion, the simultaneous evolution of dust and gas dynamics must be investigated.
In addition, the self-consistent approach enables us to consider the
effects of SI.
Since the SI becomes more effective at the dust-dense region (e.g., \citet{2007ApJ...662..613Y,2007ApJ...662..627J}), a pressure bump is expected to be a favorable
location for the growth of SI.

This paper is organized as follows.
In Sect. 2 we describe the equations and initial setup of our simulations.
In Sect. 3 we show the simulation results.
In Sect. 4 we discuss details of the mechanisms by which the pressure bump evolves.
We also discuss the relevance of our results
to the process of planetesimal formation.
Our conclusions are presented in Sect. 5.

\section{Equations and model}
\label{sec:equations}
\subsection{Equations}
We considered
the local Cartesian frame to rotate at Keplerian frequency $\Omega$ at a reference distance $r_0$ from a central star to study dust and gas motions.
The local Cartesian coordinates are $(x, y, z),$ where $x = r - r_0$,
$y$ is the tangential distance from the origin, and $z$ is the
vertical distance from the disk midplane.
We assumed an axisymmetric disk and performed
the 1D (radial) and 2D (radial-vertical) simulations.
Differences between the solid and the gas azimuthal velocities cause them to migrate radially as a result of angular momentum exchange through gas drag.
We calculated the azimuthal velocities at each grid cell in 1D or 2D simulations.
%
For the disk gas, we used isothermal hydrodynamic equations,
\begin{equation}
\label{eoc}
\frac{\partial \rho_g}{\partial t}+\nabla \left(\rho_g \mathbf{v}\right)=0,
\end{equation}
\begin{equation}
\label{isoeom}
\frac{\partial \mathbf{v}}{\partial t}+
\left(\mathbf{v} \cdot \nabla \right) \mathbf{v} =
- \frac{1}{\rho_g} \nabla P
+ 3\Omega^2 x \mathbf{\hat{x}}
- 2\mathbf{\Omega} \times \mathbf{v}
- \beta c_s \Omega \mathbf{\hat{x}}
- \frac{\epsilon}{\tau_{f}}(\mathbf{v}-\mathbf{w}'),
\end{equation}
\begin{equation}
\label{iso}
P=c_s^2 \rho_g,
\end{equation}
where $\rho_g$ is the density of the disk gas, $\mathbf{v}$ is the gas velocity, $P$ is the gas pressure, and $c_s$ is the constant sound speed, $\epsilon = \rho_d/\rho_g$ is the dust-to-gas mass ratio,
and $\tau_{f}$ is the friction (or stopping) time of solid bodies,
respectively.
The gas pressure gradient term is separated from the global one $-\beta c_s \Omega$.
We assumed that $\rho_g \propto r^q$, then the global pressure gradient term is
\begin{eqnarray}
\label{glp}
-\frac{1}{\rho_{g}}\frac{\partial P_0}{\partial r} \cong
-\frac{1}{\rho_{g}}\frac{P_0}{r}q \cong
-\frac{H}{r}q c_s \Omega=
-\beta c_s \Omega,
\end{eqnarray}
where $H=c_s/\Omega$ is the disk scale height.
In our local model, $\beta \equiv qH/r=-0.04 \ {\rm or} \ 0$ was adopted.
The value $\beta = -0.04$ is slightly lower than the value at $r=1$ AU in MMSN.
Since we set up a pressure bump with a size similar to that found by \citet{2012ApJ...747...11K}, we adopted the same value of $\beta$ as they.
On the other hand, $\beta = 0$ corresponds to the case without
a global pressure gradient.
Comparing our findings with the $\beta = 0$ result, we can define the effects of the global pressure gradient.
The last term on the right-hand side of Eq.(\ref{isoeom}) is
the back-reaction from dust onto gas.
The description $\mathbf{w}'$ is the allocated dust velocity at the grid point.
We used the cloud-in-cell (CIC) model to allocate the average dust velocity.
We assumed that the scaled
stopping timescale Stokes number $\tau_s \equiv \tau_f \Omega$ is $1.0$.
Dust particles are only characterized by the stopping timescale.
With $\tau_{s}=1$,
the dust radial drift is fastest \citep{1986Icar...67..375N}.

We included
solid materials
as super-particles.
Each super-particle consists of a large number of meter-sized boulders.
The total number of super-particles is $O(10^5 - 10^6)$ and each super-particle represents $O(10^7)$ boulders.
The equation of motion of the $i$-th
super-particle
is given by
\begin{equation}
\label{dusteom}
\frac{{\rm d} \mathbf{u}_{i}}{{\rm d} t}=
-2\mathbf{\Omega}\times\mathbf{u}_{i} + 3\Omega^2 x_{i}\mathbf{\hat{x}}
- \frac{1}{\tau_f} \left( \mathbf{u}_{i} - \mathbf{v'}_{i} \right),
\end{equation}
where $\mathbf{v'}_{i}$ is the gas velocity
at the location of the $i$-th particle,
which is interpolated using $\mathbf{v}$ at the neighboring grid points.

In simulations, the gas equations are solved by the CIP scheme \citep{1991CoPhC..66..219Y},
which is one of the methods of grid-hydrodynamics.
The dust density and velocities are allocated to the closest four (2D) or two (1D) grid points in the computational region using cloud-in-cell interpolation.
This algorithm strictly conserves angular momentum
in the combined gas-dust system.
We used a local simulation box, which means that the boundary conditions were periodic in all directions.
For the radial boundary, however, we used the shearing-box approximation, which considers Keplerian differential rotation \citep{1988AJ.....95..925W,1995ApJ...440..742H}.
These methods are similar to those used by \cite{2012ApJ...747...11K}.
\begin{table*}
\centering
\begin{tabular}{rccccc}
\hline \hline
Run & $L_x \times L_z$ & $N_x \times N_z$ & $N_p$ & $\beta$ & $T_{\rm end}$ \\
\hline

 $\beta$04-1D & 10.0$H \times -$  & 10000$\times$1 & 2.7$\times$10$^5$
              & -0.04 & 500$\Omega^{-1}$\\
 $\beta$04-2D & 7.5$H \times$0.25$H$ & 1500$\times$50
              & 1.875$\times$10$^6$
              & -0.04 & 500$\Omega^{-1}$ \\
 $\beta$00-1D & 4.0$H \times -$ & 4000$\times$1 & 1.08$\times$10$^5$
              & 0.0 & 500$\Omega^{-1}$ \\
\hline
\end{tabular}

\caption{Setup of individual runs.
$L_x, L_z$ are the radial and vertical width of computational region.
$N_x, N_z$ are  the grid numbers in radial and vertical direction.
$N_p$ is the number of super-particles.
$\beta$ is the global pressure gradient coefficient.
$T_{\rm end}$ is the total run time.
}
\label{tab:initial}
 \end{table*}
\subsection{Initial conditions}
We conducted numerical simulations with three setups.
The first was a 1D (radial) simulation with a bump.
In this run, we tested the 1D evolution of the bump in the radial direction.
The second was a 2D (radial-vertical) simulation with a bump.
These are the main results of this paper.
We investigated the pressure bump evolution and the growth of the SI at the same time.
The third was a 1D simulation without the global pressure gradient.
This setup was the same as the first, except that the global pressure gradient was included to check its effects
on the time evolution of the radial pressure bump.
Other parameters and settings for these cases are presented in Table \ref{tab:initial}.

In all types of calculations, we set the pressure bump as the initial condition.
Our pressure bump was sustained by the azimuthal velocity profile given by Eq.~(\ref{ini_vy}).
Therefore, it was steady if there was no
back-reaction from solid materials onto gas.
We note that we assumed that the pressure bump had already formed through some mechanism (e.g., MRI or baroclinic instability), and we followed the evolution of the bump without including its source mechanism in our simulation.
This approximation assumes that the timescale of the bump structure production is longer than the deformation timescale that is due to back-reaction.
The radial structure with equilibrium between dust and gas is
found by solving a time-dependent initial-value problem.

The initial pressure bump is given by
\begin{eqnarray}
\label{ini_p}
P=P_0 \left\{1+a \exp{\left(\frac{-x^2}{2h^2}\right)}\right\},
\end{eqnarray}
where $a$ and $h$ are the arbitrary parameters for the height and width of a pressure bump.
We set $a=0.2, h=\sqrt{1/4}H$ in our simulations.
The
initial azimuthal velocity profile is
\begin{eqnarray}
\label{ini_vy}
v_y=-\frac{3}{2}\Omega x +
\frac{1}{2\Omega}
\left\{\beta c_s \Omega - \frac{1}{\rho_g}\frac{aP_0 x}{h^2}
\exp \left(-\frac{x^2}{2h^2}\right)\right\},
\end{eqnarray}
which is derived from the radial force balances.
In the 2D simulation, we assumed a uniform distribution vertically
because of the narrow computational boxes here.

In all cases, the dust particles initially rotated with Keplerian velocity in the whole computational region.
The initial particle distribution was uniform in our computational domain for simplicity,
and the initial dust-to-gas mass ratio was $\epsilon=0.1$.
The initial dust-to-gas mass ratio $\epsilon = 0.1$ is higher than $0.01,$ which is the column dust-to-gas mass ratio in MMSN.
Since we considered the solid materials to be meter-sized boulders, the scale height of the solid materials is more than $10$ times smaller than the gas scale height \citep{2007Icar..192..588Y}.
Therefore the dust-to-gas ratio is $\sim 0.1$ at the disk midplane.
After a few frictional timescales, the dust particles attain terminal drift velocity given by the Nakagawa-Sekiya-Hayashi equilibrium solution \citep{1986Icar...67..375N}, except near the region of the bump, which is given by
\begin{eqnarray}
\label{NSHvr}
u_{r,\mathrm{NSH}}=\frac{\tau_f \Omega}{\left(\tau_f \Omega\right)^2+\left(\epsilon+1\right)^2}\beta c_s.
\end{eqnarray}

To ensure that radial migration caused by gas drag provides a continuous supply of dust particles to the bump until run time $T_{\rm end} = 500$, we set a sufficient radial width for the computational box.
Since the radial drift velocity was $u_{r,\mathrm{NSH}}= -0.018c_s$ in our 1D simulation, the drift length is $9H$.
In our 2D simulation, the dust radial velocity decreases by $\sim 40\%$ because of the SI \citep{2007ApJ...662..627J}, and the drift length is $\sim 5.4H$.
Then the box size in the radial direction is $L_x=10H$ (1D simulations), $7.5H$ (2D simulation), which exceeds the drift length.
The size of the vertical box is $L_z=0.25H,$ which is small enough compared to the disk scale height of the gas.
\section{Results}
\subsection{1D dust-density evolution}
\begin{figure}
\resizebox{\hsize}{!}{\includegraphics{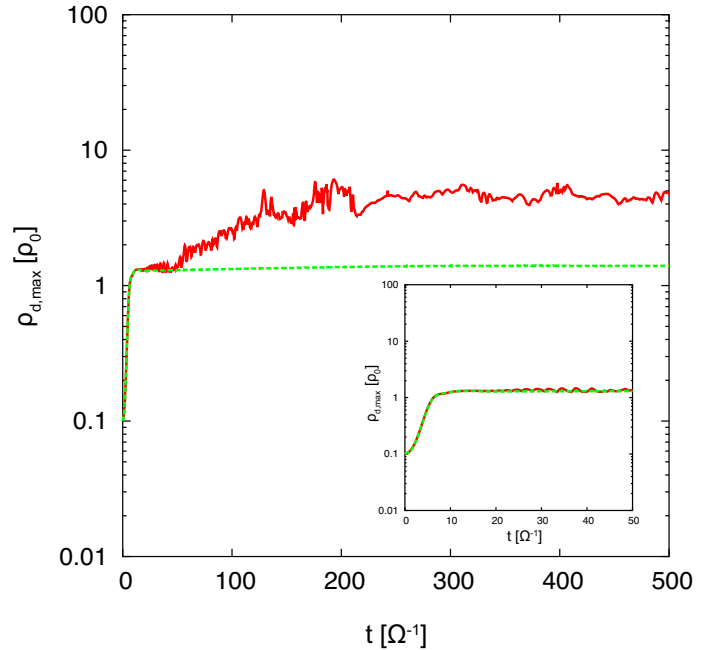}}
\caption{
Time evolution of the highest dust density in run $\beta$04-2D (2D w/ bump; red line) and $\beta$04-1D (1D w/ bump; green dashed line).
All the lines represent the dust density in the cell with the highest density of the whole computational region, normalized by the background gas density $\rho_0$.
The initial dust accumulation due to the bump structure is almost the same in the two runs (see inset).
}
\label{fig1}
\end{figure}
\begin{figure*}
\centering
\includegraphics[width=17cm]{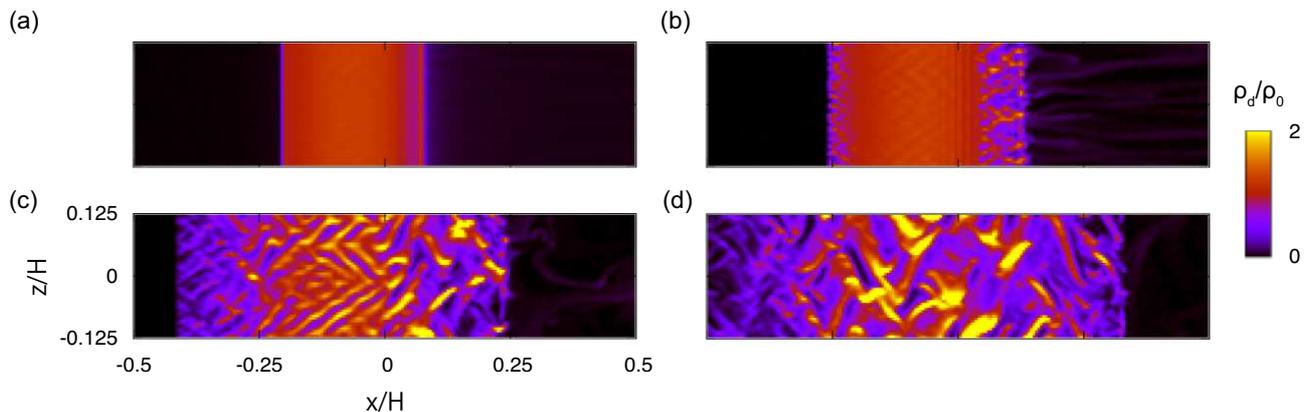}
\caption{
Dust-density pattern obtained from run $\beta$04-2D (2D w/ bump).
Each panel represents a snapshot at (a) $t\Omega=50$, (b) $100$, (c) $250$, and (d) $500$.
Dust densities increase from black (zero density) to bright white ($\rho_d = 2 \rho_0$).
 Initially, the pressure bump concentrates the dust particles,
with the dust-to-gas mass ratio being nearly unity (panel (a)).
Subsequent panels show the growth of SI in this dust-dense region.
We note that the region near the pressure bump is zoomed-in and the whole computational domain is radially wider.
}
\label{fig2}
\end{figure*}

First, we describe the 1D view of the pressure bump evolution.
The pressure bump halts inward, accreting dust particles at the point of highest pressure.
If there is no back-reaction
from dust to gas, the pressure bump is steady and the dust density increases until the gravitational collapse of solid components occurs,
which is the expected result
in a test-particle approach.
When the dust drag is taken into account, the pressure bump is deformed by the angular momentum exchange and the dust particles are not concentrated at the point, but are distributed over a radial range.
The timescale of the entire deformation of the pressure bump (i.e., there is no longer a super-Keplerian region) is $\sim 150\Omega^{-1}$.
Since the radial dust accumulation causes the deformation of the pressure bump, this timescale depends on the global pressure gradient, the initial dust-to-gas ratio, and the stopping time of the particles.
The radial width of the dust-dense region is about half that of the pressure bump.
We assumed that the radial width of the pressure bump is $\sim 2H$.
The width of the dust-dense region is $\sim H$.

Figure \ref{fig1} shows the time evolution of the highest dust density, which is found
in the cell with
the highest density of
the whole computational region.
In Fig. \ref{fig1} the green line represents the results of the run $\beta$04-1D.
The highest dust density increases until $t\Omega \sim 10$ because the dust is trapped in
the pressure bump.
At $t\Omega \gtrsim 10$, the density stops increasing and the saturated dust density is $\sim \rho_0$.
The mechanism of this saturation is the same as with the 2D simulation (run $\beta$04-2D) and
the simpler 1D simulation without global pressure (run $\beta$00-1D), which
we discuss in subsequent sections.

The dust-dense region has a dust-to-gas mass ratio on the order of unity.
Although a much higher dust-to-gas ratio is required for a simple Toomre-type self-gravitational instability to occur, \cite{2007ApJ...662..627J} suggested that when two-dimensionality is introduced, the SI causes strong clumping of dust particles in such a dust-dense region.
That is, a pressure bump is one of the most probable regions where the SI is excited and dust clumps are formed.
In our simulations to be described in the next subsection, however, the behavior of SI is found to be different from previously studied cases because of the pressure bump deformation.
\subsection{2D dust-density evolution}
Based on 2D simulation results, we confirm that the formation of the dust-dense region is due to dust trapping and the growth of SI in the dust-dense region.
In Fig. \ref{fig1} the red line shows the time evolution of the highest dust density in the whole computational region from the 2D simulation.
Until $t\Omega \sim 50$, the highiest dust density evolves as in the 1D case.
Then, the red line surpasses the green one, indicating that the linear growth of SI has begun.
The dust concentration is saturated by $t\Omega \sim 200$, with the highest dust density converging at $\sim 5\rho_0$.

Figure \ref{fig2} shows snapshots of the dust-density profile at the bump.
In panel (a, b) of Fig. \ref{fig2}, which are the snapshots at $t\Omega = 50, 100$, the dust-dense region is formed by the radial pressure bump.
The region has a nearly unity dust-to-gas mass ratio as in the 1D simulation.
In panel (c), the density spatial pattern shows the linear growth of the SI.
In panel (d), the saturated state of the instability is shown.
There are some large dust clumps with densities higher than $2\rho_0$.

Based on Figs. \ref{fig1} and \ref{fig2}, we confirm that SI forms dust clumps and that the highest dust density of these clumps is $\sim 5 \rho_0$ in the dust-dense region.
On the other hand, \citet{2007ApJ...662..627J} reported the highest dust density from clumping created by SI to be $\sim 100\rho_0$ for $\tau_{s}=1.0$ and $\epsilon =1.0$.
\citet{2007ApJ...662..627J} employed 
an initially uniform dust and gas density of $\sim \rho_0$ that was distributed throughout the whole computational domain.
Our highest density is quite low compared with the value obtained from the other case, even though the dust-to-gas ratio that excites the SI is equally high $\sim 1$.
This means that the SI with the spatial scale observed in previous studies is inefficient at the pressure bump.
The reason for this weaker clumping in our case is discussed in subsequent sections.

\begin{figure}
\resizebox{\hsize}{!}{\includegraphics{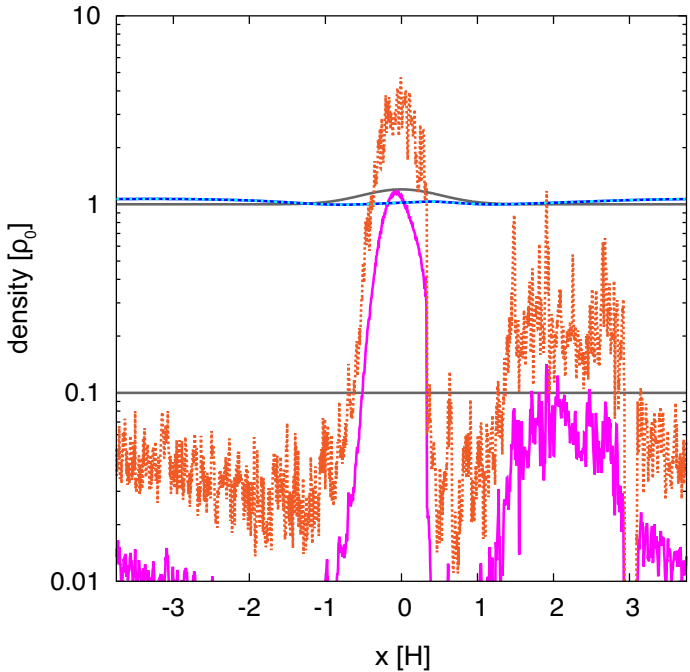}}
\caption{
Dust and gas density radial profile from run $\beta$04-2D (2D w/ bump).
Each line is a snapshot at $t\Omega=500$.
Gray solid lines show the initial profile of gas (upper line) and dust (lower line) densities.
The orange dotted line and the magenta solid lines show the highest and vertically averaged dust densities at each $x$.
The vertically averaged dust density stays at $\sim 1$ as in the 1D run, while the upward deviation of the orange line from the pink one shows the growth of SI in the 2D case.
The cyan solid and blue dotted lines showing the gas densities (highest and vertically averaged, respectively) show small differences that imply that the gas density has a nearly uniform distribution in the vertical direction.
}
\label{fig3}
\end{figure}

Figure \ref{fig3} shows a snapshot of the vertically averaged and
highest values of dust and gas densities from the 2D result at $t\Omega =500$.
The orange and magenta
lines show the highest and vertically averaged dust density at each radial location, respectively.
The deviation of the orange line from the magenta line
shows that the SI is excited throughout the dust-dense region.
The average dust density at the pressure bump is $\sim \rho_0$ and is almost the same as the 1D results.
This means that the basic dust-density radial structure is due to the 1D dynamics and that SI adds modulation along the vertical direction.

The blue and the cyan lines in Fig. \ref{fig3} show the vertically averaged and highest gas densities at each radial location $x$,
respectively.
There is a small difference between the two, implying that SI does not result in a strong vertical variation of gas density.

%
\subsection{Evolution of the gas pressure profile}
\label{sec:evol-gas-press}
In our simulations, we followed the time evolution of the local gas pressure, assuming that the global pressure gradient was constant.
Here we study
the evolution of the total pressure gradient force in the dust-dense region.
The total gas pressure is a sum of the assumed constant global pressure Eq.~(\ref{glp}) and the local gas pressure $P$ calculated in our simulations.

\begin{figure*}
\centering
\includegraphics[width=17cm]{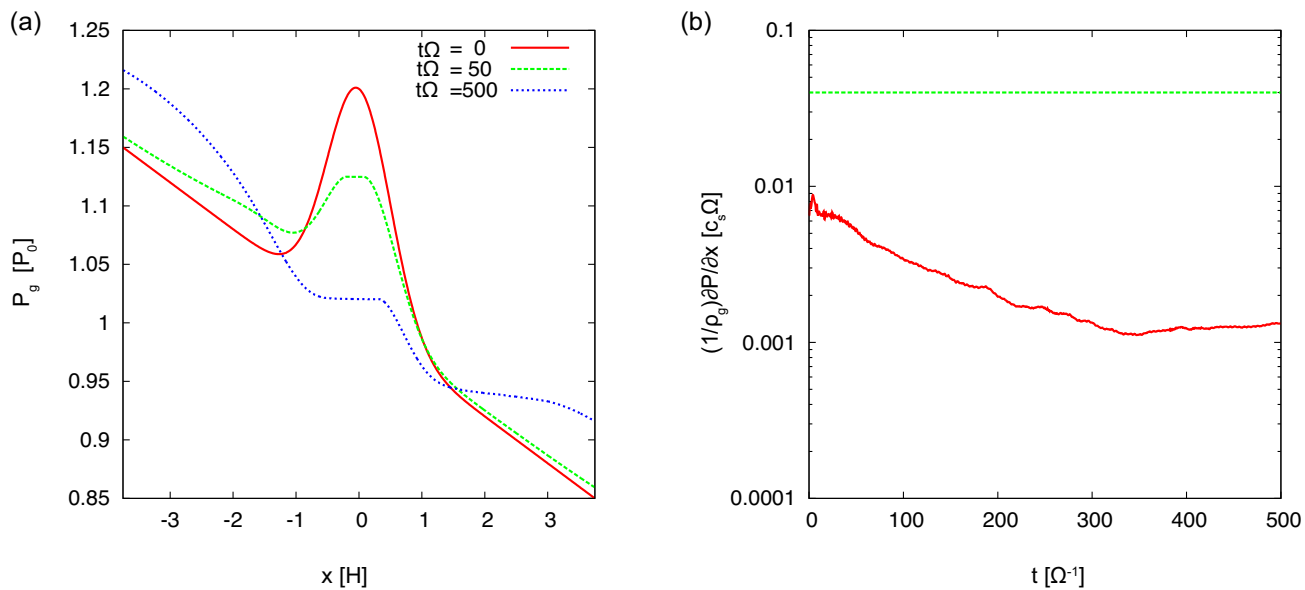}
\caption{
(a) Vertically averaged gas pressure profile and (b) time evolution of the radial pressure gradient force averaged spatially in the dust-dense region from run $\beta$04-2D (2-D w/ bump).
As the dust particles accumulate at the bump, the total gas pressure profile is flattened.
Then the pressure gradient force at the bump converges to $\sim 0.0013$, which is lower than the global pressure gradient $\beta=-0.04$ set initially (green dashed line).
}
\label{fig4}
 \end{figure*}

Expressions for the total gas pressure and its gradient
are
\begin{eqnarray}
\label{totallpressure}
P_{\mathrm{total}} &=& P + \beta c_s \Omega \rho_0 x, \\
-\frac{1}{\rho_g}\frac{\partial P_{\mathrm{total}}}{\partial x} &\cong& -\frac{1}{\rho_g}\frac{\partial P}{\partial x} - \beta c_s \Omega.
\end{eqnarray}
Figure \ref{fig4}(a) shows snapshots of the total pressure at $t\Omega =0, 50, 500$ from the 2D case.
At the pressure bump, the total pressure profile tends to become flat over time, and finally at $t\Omega=500,$ the whole dust-dense region has an almost zero total pressure gradient.
In Fig. \ref{fig4}(b), to confirm pressure flattening, we check the evolution of the pressure gradient force in the dust dense region ($-0.2H < x < 0.1H$).
The spatially averaged pressure gradient converges to $\sim 0.0013,$ while the initial value of the global pressure gradient is $\mid \beta \mid=0.04$.
This means that the pressure gradient force at the dust-dense region is $\sim 10-100$ times lower than the value is initially set for the region outside the bump.
As we discuss below, the significantly lower pressure gradient plays an important role in the size of the dense particle clumps formed by SI.

\begin{figure*}
\centering
\includegraphics[width=17cm]{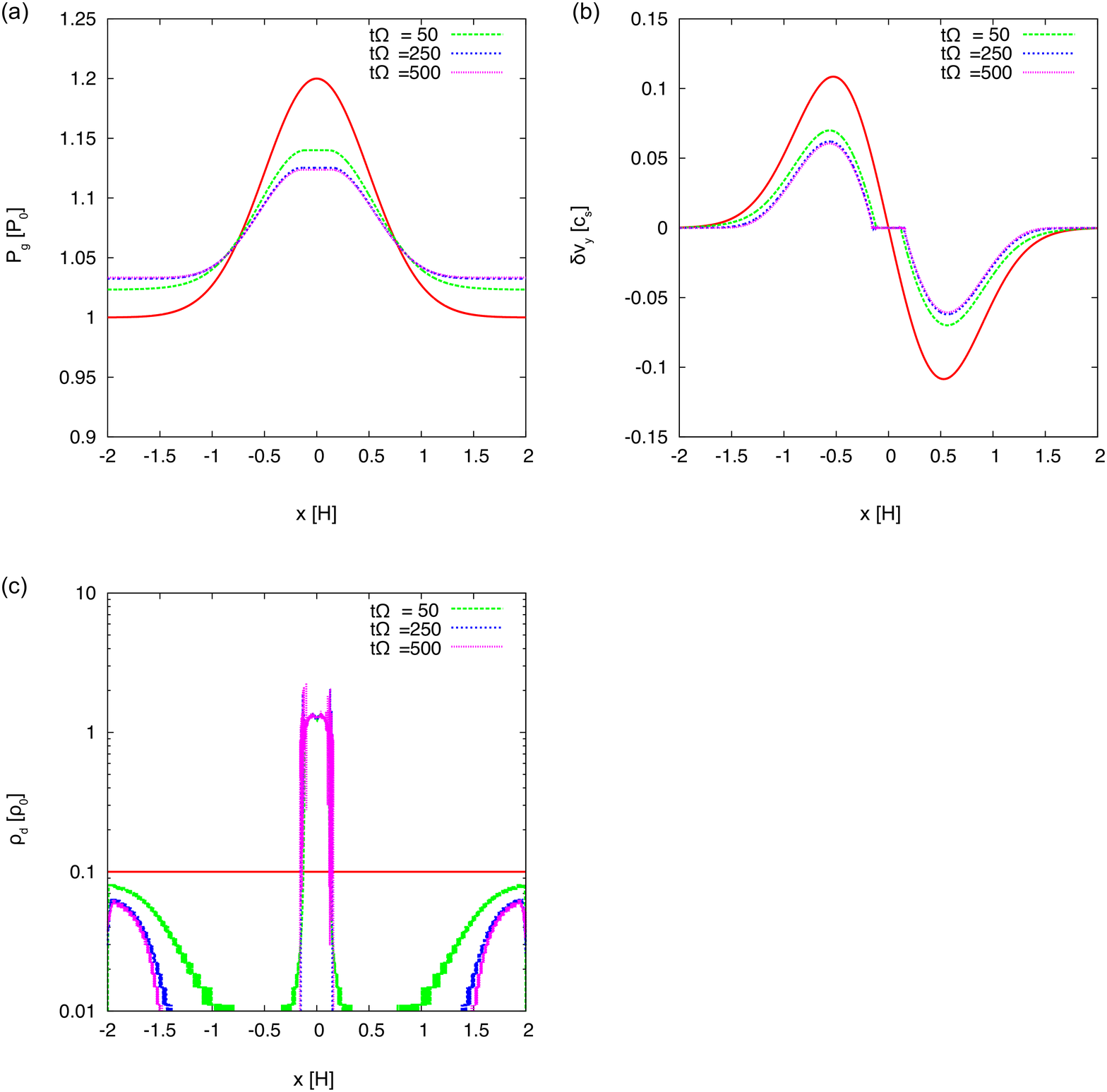}
\caption{
Results from the model without the global pressure support (run $\beta$00-1D).
(a) gas pressure, (b) gas azimuthal velocity (the deviation from the Keplerian rotation), and (c) dust density. The essential physics of the dust-enriched region can be learned from these simple simulation results.
}
\label{fig5}
\end{figure*}

To investigate a simpler case without the global pressure gradient, we conducted a 1D simulation with $\beta=0$.
In this case, there was no
accumulation of particles coming from regions outside the pressure bump,
but dust was concentrated in
the pressure bump toward the point of the pressure maximum.
Figure \ref{fig5} shows the snapshots of the gas profile evolution ($t\Omega = 50, 250, 500$).
Panel (a, b) shows the gas pressure profile and the gas azimuthal velocity profile (deviation from Keplerian rotation).
Panel (c) shows the dust density profile.
The
system achieves steady-state
by $t\Omega \sim 250$.
In the dust-dense region, the dust density is $\sim \rho_0$, the gas azimuthal velocity is $\sim 0$, and the gas pressure profile is flattened.
That is, the velocity distribution of gas is also forced to follow a Kepler rotation
that is due to strong drag from dust particles in the area with a dust-to-gas ratio $\sim 1$.
This clearly depicts the essential nature of what is seen in the 2D or 1D simulations with the global pressure gradient.
The only difference is that in the simulations with a global pressure gradient, the completely flattened pressure profile (the dust-dense region) expands outward as a result of the continuous inward radial drift of dust.

\section{Discussion}
\subsection{Schematic of the pressure bump evolution}
In this subsection we discuss the mechanism for dust and gas density evolution at the radial pressure bump.
First, we describe a 1D picture of the evolution in the radial direction.
The 1D evolution can be separated into two parts.
One is the saturation of the dust density that is due to the pressure gradient evolution.
The second is the deformation of the entire pressure profile as a result of the continuous
trapping of particles at the outer edge
of the pressure bump from the outer part of the disk.

The saturation of dust density occurs when the dust and gas achieve equilibrium near the center of the radial pressure bump.
Initially, the dust density grows faster than the gas density evolution.
This is because the
inward drift velocity is
faster than the outward flow velocity of gas when the dust-to-gas mass ratio is lower than unity.
As the dust density increases near the center of the pressure bump, the dust-to-gas mass ratio becomes $\sim 1$.
Then the gas radial velocity becomes similar to the dust radial drift velocity, meaning that the timescale of the gas density evolution is equalized with the timescale of the dust density increase.
The outward gas flow flattens the pressure bump and continues until the total pressure gradient force becomes $\sim 0$.
Upon reaching the state where the pressure gradient force is $\sim 0$, the dust and gas have almost zero radial velocity and achieve  steady-state. The equilibrium is such that both dust and gas are mostly in the Kepler rotation state.
Since additional density concentration cannot be expected, the highest dust concentration saturates at a dust-to-gas mass ratio
of $\sim 1$.

Regarding the second point of the density evolution mechanism at the radial pressure bump,
the entire pressure bump deformation is brought about by the continuous dust trapping at
the outer edge of the pressure bump.
When there is continuous dust trapping,
the gas component continuously leaks outward.
This outward-flowing mechanism continues until the dust trapping
ceases or the pressure bump is destroyed entirely.

The simpler 1D calculation without global pressure gradient helps us understand the bump evolution very clearly
since the final stage does not change in time and thus the interpretation becomes easier.
Since there is no global pressure gradient $\beta = 0$, there is no dust trapping
from outside the pressure bump, but the dust particles that initially existed
inside the pressure bump accumulate at the super- or sub-Keplerian transition point.
The results shown in Fig. \ref{fig5} indicate that the highest dust density of $\rho_{\mathrm{d, max}} \sim 1.4\rho_0$, which is the same as for $\beta=-0.04$, is obtained in the region where the gas pressure gradient is nearly $0$.
The flattened pressure profile in the dust-dense region is in a steady state of equilibrium.
That is, the dust concentration stops when both gas and dust achieve
Kepler rotation.
A dust-to-gas mass ratio of $\sim 1$ corresponds to a scenario
where the back-reaction
becomes strong enough to force the gas into Kepler rotation.
The size of the dust-dense region is $\sim 0.3H$.
This is determined by the initial total mass of the particles in the pressure bump.

Results of the 2D simulation are explained by the 1D picture with the addition of
vertical modulation due to SI.
After the formation of a dust-dense region with the dust-to-gas ratio of $\sim 1$ that is due to the 1D dynamics, SI is excited because of the high dust density.
We note that the pressure gradient in the dust-dense region is significantly
reduced compared to
initial conditions, which affects the
properties of SI that could be
excited in this environment, as discussed below.


\subsection{Properties of SI in the dust-dense region}
In this section, we discuss the detailed properties of SI excited in the dust-dense region formed by the radial pressure bump.
The dust-dense region can be regarded as the initial conditions of SI.
We found that in the dust dense region, (i) the dust-to-gas mass ration is $\sim 1$ and (ii) $\eta$ is reduced by $10-100$ times from the nominal value of MMSN.

The relatively high dust-to-gas ratio is favorable for SI \citep{2007ApJ...662..613Y,Carrera2015}.
Especially when the stopping timescale $\tau_s$ is shorter than $\sim 1$,
the pressure bump may be important for SI to occur.
The bump deformation mechanism does not strongly
depend on the frictional timescale.
If we neglect gas turbulence, the bump deformation process becomes slower, but the saturation mechanism of dust density does not change even for millimeter-sized particles.
While the dust-to-gas mass ratio also becomes $\sim 1$ for small particles in the dust-dense region,
the growth rate of SI rises drastically at a dust-to-gas mass ratio $\ga$ 1 \citep{2007ApJ...662..627J}.
Therefore, the pressure bump may be a good location for planetesimal formation, even for small dust particles that have a short stopping time of $\tau_s \sim 0.003$ corresponding to the mm-sized object at $r=1$ AU in MMSN.

The evolution of pressure gradient, however, 
affects the properties of SI at the pressure bump.
The spatial scales of SI are normalized by $\eta r$, where $\eta$ is another expression of the global pressure gradient $\eta = -H\beta/(2r)$ \citep{2005ApJ...620..459Y}.
As seen in Sect. \ref{sec:evol-gas-press}, the total pressure gradient at the pressure bump becomes $\sim 10-100$ times lower than the initial global value.
Accordingly, the spatial scale of SI also becomes $\sim 10-100$ times smaller than the value of previous studies.
This means that SI with the spatial scale observed in previous studies has a reduced growth rate.
We did not observe dense particle clumps with a size similar to those found in \citet{2007ApJ...662..627J}.
The most unstable scale of dense particle clumps at the pressure bump predicted by our results
is also $10-100$ times smaller than that given by \citet{2007ApJ...662..627J}.
\citet{2012A&A...537A.125J} conducted simulations including many effects (e.g., vertical gravity, particle self-gravity, and particle collision) to find that the SI forms gravitationally bound clumps corresponding to $\sim 100-1000$ km sized bodies with a normal MMSN value of $\beta \sim 0.1$.
If the SI occurs in the same way as reported by \citet{2012A&A...537A.125J}
at the pressure bump, the predicted clump size is $10-100$ times smaller than their results.
The predicted size is $\sim 1-100$ km sized bodies,
which is similar to the classically predicted planetesimal size.

We note that the spatial resolution of our calculations is insufficient to resolve this small-scale
SI in the dust-dense region.
We assumed that the width of the radial pressure bump is similar in size to the gas pressure scale height.
To ensure a sufficient width for the dust drift and accumulation, our computational box has $\sim 10$ scale heights in the radial direction.
However, the predicted
short wavelength of the SI in the dust-dense region requires higher resolution.
Since the wavelength of the SI becomes short but the growth rate does not change even if the global pressure gradient becomes small,
a simple prediction is that the highest dust density is similar to the results of the case with $\tau_s=1.0, \epsilon =1.0$ in
\citet{2007ApJ...662..627J}, which is almost the same setting as with the dust-dense region formed by the pressure bump.
Thereby, it is not clear in our simulations whether the SI with a short wavelength does occur.

\subsection{Future works on planetesimal formation at the radial pressure bump}
Our calculations did not take the Kelvin-Helmholtz instability (KHI) by radial shear in azimuthal velocity and vertical shear in radial or azimuthal velocity into account.
\citet{2007ApJ...662..627J} and \citet{2010ApJ...722.1437B} found that SI  dominantes the KHIs.
In general, small-scale instability can be damped by other perturbations.
At the radial boundary of the dust-dense region, there must be a shear stronger than the Keplerian one
to cause strong KHI.
The investigation of the effect of KHI is left for future work.

Dust self-gravity and vertical gravity of the host star are other important physics neglected in this study.
\citet{2010ApJ...722.1437B} and \citet{2012A&A...537A.125J} included them and confirmed that gravitationally bound clumps are formed by SI at the midplane dust layer.
This may also be applied to the pressure bump.

We set the radial pressure bump as the initial condition of our calculations.
Various formation mechanisms for pressure bumps have been proposed (e.g., MRI zonal flows, edges of the MRI dead zone, and a snow line).
These mechanisms and the global disk evolution affect the bump deformation process.
If the bump restoration or restructuring process is faster than the destruction process we discussed here, the radial pressure bump could accumulate many more dust particles.

\section{Summary}
We numerically studied the dust and gas density evolution at a radial pressure bump in a protoplanetary disk.
Disk gas dynamics was treated with grid hydrodynamics and dust components with the particle-in-cell scheme.
Computational regions were approximated by the local shearing-box model.
The gas drag force to dust and its back-reaction to gas were consistently
included in our simulations.

Our conclusions are summarized
as follows:
\begin{itemize}
 \item The pressure bump accumulated dust particles, but the outward flow of gas created by the back-reaction smoothed it out after the dust-to-gas mass ratio reached $\sim 1$.
However, this only occurred if
bump restoration or restructuring was slower than the destruction process.

 \item The pressure bump created a long-lived dust-dense region
       where the vertically averaged dust-to-gas mass ratio was $\sim 1$.
       When the inward drift of dust
       continued to deliver dust particles to the outer edge of the pressure bump, the flattened pressure gradient region
       in the dust-dense region expanded outward until the pressure bump was completely destroyed.

 \item As a result of bump destruction, direct self-gravitational instability at the site of the
       pressure bump was inhibited.

 \item The gas flow in pressure bumps was modulated to be close to the Keplerian one and the deviation from Keplerian velocity ($\eta$) was $\sim 10-100$ times smaller than the global value.
       Because the SI wavelength is scaled by $\eta$, the clumps formed by SI, even if the SI occurs, should
       become $1 - 100$ km sized bodies, which are much smaller than the previously proposed size ($\sim 100-1000$ km).
       This means that the SI forms planetesimals with sizes similar to those formed by the self-gravitational instability of a thin dust layer \citep{1972epcf.book.....S,1973ApJ...183.1051G} at the radial pressure bump.
\end{itemize}

\begin{acknowledgements}
 We thank an anonymous referee for detailed comments.
 We are grateful to Taishi Nakamoto, Taku Takeuchi, Satoshi Okuzumi and Akimasa Kataoka for comments.
 This research was supported by a grant for JSPS (23103005) Grant-in-aid for Scientific Research on Innovative Areas.
\end{acknowledgements}
\bibliographystyle{aa}
\bibliography{ref}
\end{document}